\documentclass[aps,prl,showpacs,floats,twocolumn,amsfonts,amssymb,unsortedaddress,floatfix]{revtex4}
\input epsf
\usepackage{bm}
\usepackage{amsmath}
\usepackage{graphicx}

\begin{document}

\renewcommand{\bottomfraction}{0.7}
\renewcommand{\topfraction}{0.7}
\renewcommand{\textfraction}{0.2}
\renewcommand{\floatpagefraction}{0.7}
\renewcommand{\thesection}{\arabic{section}}
\addtolength{\topmargin}{10pt}

\title{Ground state lost but degeneracy found: \\
 The effective thermodynamics of artificial spin ice} 
\author{Cristiano Nisoli$^{*}$, R. F. Wang$^{*}$, Jie Li$^{*}$, William
  F. McConville$^{*}$, Paul E. Lammert$^{*}$, Peter Schiffer$^{*}$ and Vincent H. Crespi$^{*\dagger}$}
\affiliation{$^{*}$Department of Physics and Materials Research Institute \\
$^{\dagger}$Department of Materials Science and Engineering \\ The
  Pennsylvania State University, University Park, PA 16802}
 
\date{\today}
\begin{abstract} 
We analyze the rotational demagnetization of artificial spin ice, a recently realized array of nanoscale single-domain ferromagnetic islands. Demagnetization does not anneal  this model system into its anti-ferromagnetic ground state: the moments have a static disordered configuration similar to the frozen state of the spin ice materials.  We demonstrate that this athermal system has an effective extensive degeneracy and we introduce a formalism  that can predict the populations of local states in this ice-like system with no adjustable parameters.
\end{abstract}

\date{\today}


\maketitle

\begin{figure}[t!!]
\begin{center}
\epsfxsize 3 in
\epsfbox{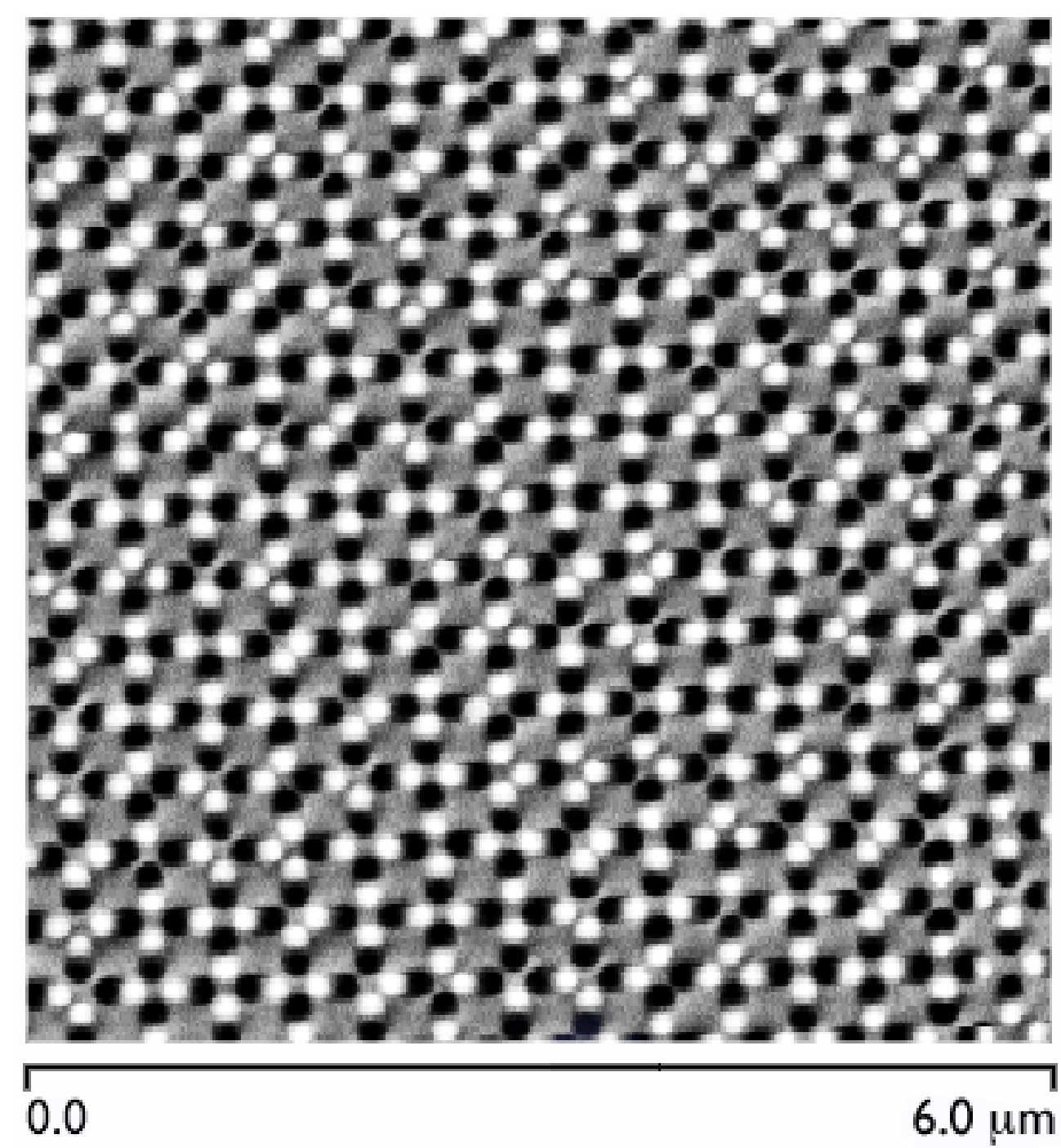}\vspace{-4 mm}
\end{center}
\caption{A representative magnetic force  microscope image of the experimental system. The single domain character of islands is indicated by the division of each island into black and whites halves~\cite{Wang}. } 
\vspace{-2 mm}
\label{MFMExperiment}
\end{figure}
\begin{figure}[t!!]
\begin{center}
\vspace{5 mm}
\epsfxsize 3 in
\epsfbox{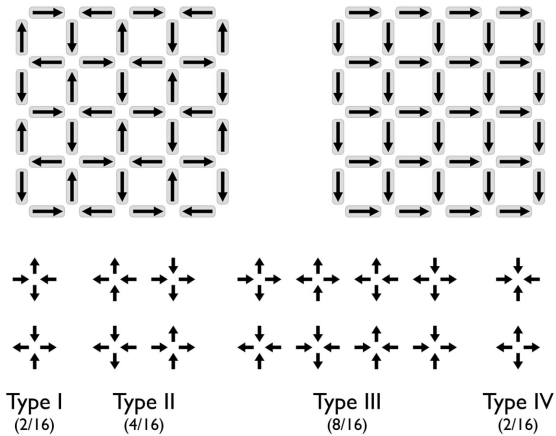}\vspace{-4 mm}
\end{center}
\vspace{-2 mm}
\caption{Top: The sixteen possible vertices of artificial spin ice and their multiplicities. Bottom: The demagnetized tiling of the ground state (left) and a fully polarized tiling of type-II vertices (right).} 
\label{structure}
\end{figure}

Disordered states are complex and often do not reveal themselves completely to experiment.  In neural networks, structural glasses, economic models and countless other systems, disorder is often associated with frustration, a competition between interactions, not all of which can be satisfied.  Even perfectly regular lattices can exhibit frustration if the interactions on the lattice have a fundamental geometrical incompatibility, such as for antiferromagnetic interactions around a three-fold loop. Such geometrical frustration governs the exotic ground states of certain spin systems~\cite{Ramirez1,Moessner} such as spin ice, wherein the spin interactions mimic the frustration of proton positions in water ice and so produce a degenerate ground state with an extensive zero-temperature entropy~\cite{Pauling, Harris, Ramirez2, Bramwell}. Recently, Wang {\it et al.}  \cite{Wang} have fabricated {\it artificial spin ice}\/: a two-dimensional array of elongated single-domain permalloy islands whose shape anisotropy defines Ising-like spins arranged along the sides of a regular square lattice, as in Figs.~\ref{MFMExperiment} and \ref{structure}.  The island-island interactions in such a lattice can be engineered to display frustration and the magnetic state of every island can be revealed by scanning probe microscopy. Frustration-induced disorder on a regular lattice, fully resolved experimentally, provides a powerful model system.  As reported in \cite{Wang}, demagnetized artificial spin ice shows short-range ice-like correlations and no long-range correlations, with consistent vertex populations across experimental runs. A recent paper~\cite{Moller} examined this system in the context of the two-dimensional vertex models of spin ice~\cite{Baxter,LW,Lieb}. 

The six pairwise island-island interactions at a given vertex of Fig.~\ref{structure} cannot be simultaneously satisfied.  However, due to a difference in the pairwise island interaction energies for neighbors at $\pi$ and $\pi/2$ angles to each other around a given vertex, this array actually has a single ground state, unique up to a global spin flip, that carries no macroscopic moment.   In \cite{Wang}, the system was demagnetized by first fully polarizing it in a large external field and then gradually decreasing the field in an alternating stepwise fashion while rapidly rotating the sample, all at room temperature~\cite{protocol}.  Although the resulting magnetic states had only a small residual moment and a much lower energy than the fully polarized starting configuration, the unique ground state was never reached, nor even closely approached.  Here, we analyze experimental data on artificial spin ice~\cite{Wang} and demonstrate that the demagnetization protocol generated a well-defined (albeit {\it not} thermally equilibrated) disordered state which restores the macroscopic degeneracy, but on a hidden manifold. The observed vertex populations can then be accounted for as the maximally likely outcome within a stochastic model of demagnetization.
Since the disordered configurations of island moments have very small net magnetization both locally and globally~\cite{Wang}, long-range magnetic interactions are weak and do not sum coherently. The dominant interactions within the array connect the nearest-neighbor islands that comprise a given vertex.  Hence we use a 16-vertex model~\cite{Baxter, LW, Lieb}, with the vertices of Fig.~\ref{structure}: the mutual magnetostatic energy of the four islands comprising a given vertex is written as $E_{\mathrm{\scriptscriptstyle I}}$, $E_{\mathrm{\scriptscriptstyle     II}}$, $E_{\mathrm{\scriptscriptstyle III}}$ or $E_{\mathrm{\scriptscriptstyle IV}}$ and is denoted as a ``vertex energy''.  All vertex energies are calculated with fully relaxed micromagnetic simulations of isolated vertices. For simplicity, we set $E_{\mathrm{\scriptscriptstyle I}}=0$.  For the arrays studied, the vertex energies maintain the relative order
$E_{\mathrm{\scriptscriptstyle I}}< E_{\mathrm{\scriptscriptstyle II}}
< E_{\mathrm{\scriptscriptstyle III}} < E_{\mathrm{\scriptscriptstyle
    IV}}$ with a typical energy scale $E_{\mathrm{\scriptscriptstyle
    II}} - E_{\mathrm{\scriptscriptstyle I}} \sim 2~\times~10^5$ K
($\sim~3$ attojoules) at a lattice constant of $320$ nm~\cite{OtherTypes}.  This large energy scale suppresses thermal fluctuations at room temperature, as does the large energy barrier associated with the shape anisotropy of the islands, so the spin configurations obtained after demagnetization are static.

\begin{figure}[!t]
\epsfxsize 3.3 in
\epsfbox{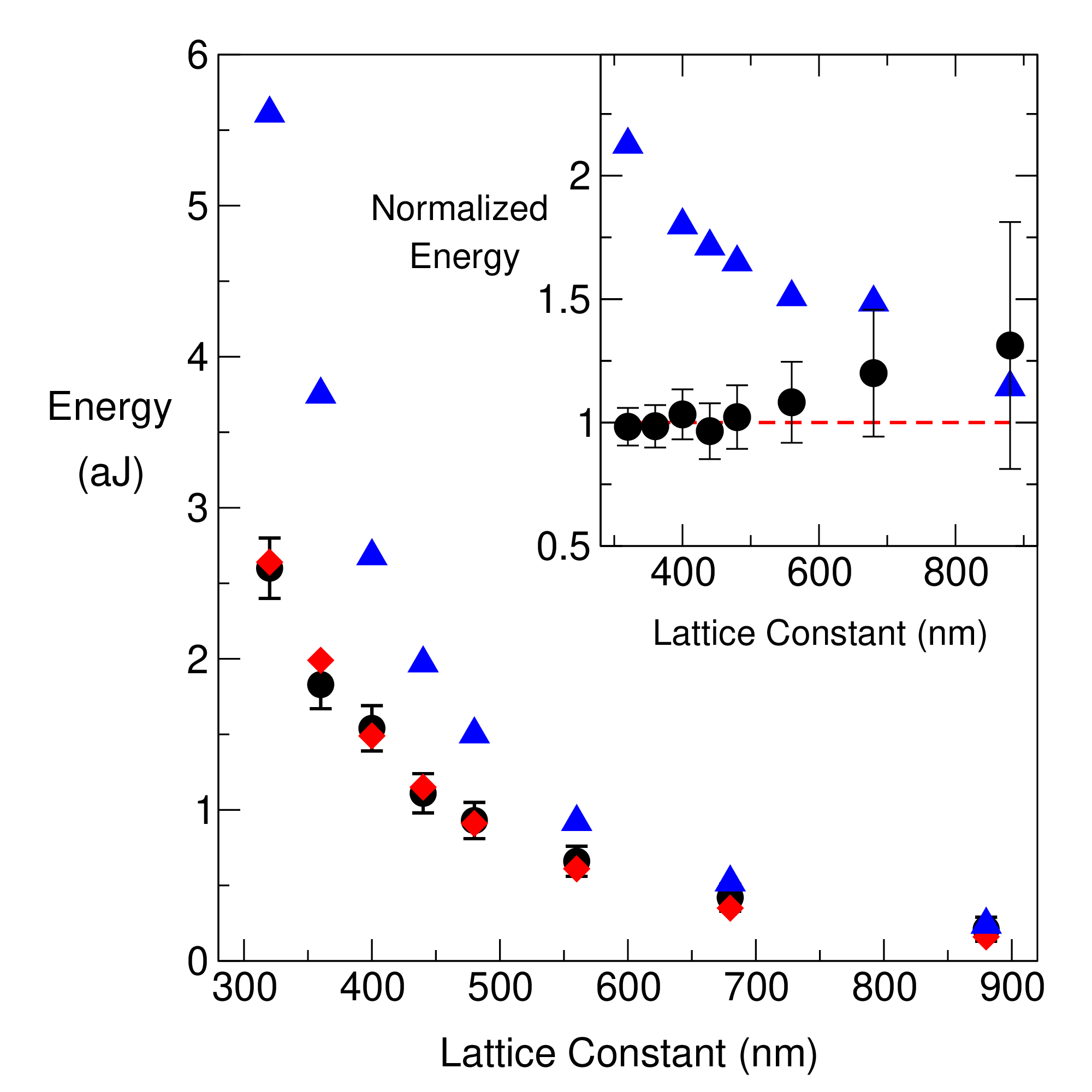}\vspace{-2 mm}
\caption{The specific vertex energy $\overline{E}$ of demagnetized lattices (dots) compared to the calculated energy of a type-II vertex ($E_{\mathrm{\scriptscriptstyle II}}$, diamonds) and the   specific energy of a purely random tiling (triangles) for a range of array lattice constants. Error bars incorporate the measured standard deviations of $n_{\mathrm{\scriptscriptstyle I}},   n_{\mathrm{\scriptscriptstyle II}}, n_{\mathrm{\scriptscriptstyle III}}, n_{\mathrm{\scriptscriptstyle IV}}$ across multiple arrays, plus a 5\% uncertainty in the micromagnetics. The inset shows the specific vertex energy and the random tiling energy, normalized to the pure type-II vertex energy at each lattice constant. One attojoule is $7.25 \times 10^4$ Kelvins}
\vspace{-2 mm}
\label{energy}
\end{figure}

If the fractional populations of the four vertex types are denoted
$n_{\mathrm{\scriptscriptstyle I}}$, $n_{\mathrm{\scriptscriptstyle
    II}}$, $n_{\mathrm{\scriptscriptstyle III}}$,
$n_{\mathrm{\scriptscriptstyle IV}}$, then the specific vertex energy
$\overline{E} = E_{\mathrm{\scriptscriptstyle I}}
n_{\mathrm{\scriptscriptstyle I}} +E_{\mathrm{\scriptscriptstyle
    II}}n_{\mathrm{\scriptscriptstyle II}}
+E_{\mathrm{\scriptscriptstyle III}}n_{\mathrm{\scriptscriptstyle
    III}} +E_{\mathrm{\scriptscriptstyle
    IV}}n_{\mathrm{\scriptscriptstyle IV}}$.  The population fractions have been extracted by magnetic force microscopy on demagnetized arrays. The measured specific vertex energy, plotted in Fig.~\ref{energy}, closely tracks that of a pure type-II tiling: $\overline{E}\approx E_{\mathrm{\scriptscriptstyle II}}$.  Even samples subjected to an abbreviated anneal protocol, whose residual magnetization is 60\% of the saturation value, obey this energetic constraint, as does the original fully polarized sample.  This striking relationship does not arise from simple random averaging-- the same plot also shows the average vertex energy of a random tiling, wherein each vertex type occurs according to its multiplicity. Apparently, the rotational demagnetization protocol, which begins with an array polarized into a uniform type-II tiling, does not significantly reduce the vertex energy; rather, it moves the configuration about within that energy manifold, reducing the total magnetostatic energy only through a reduction in the long-range demagnetization fields~\cite{couple}.  We take this experimental observation as given, and explore its consequences~\cite{ansatz}.  Population fractions for different demagnetized samples at the same lattice constant are very similar, presumably reflecting some well-defined underlying state.  Individual islands for samples with lattice constants of 320 and 560 nm show no significant correlations in moment orientation between successive demagnetization runs, so the moments are not pinned by static structural disorder in island shape. All samples analyzed below had a residual magnetization $M_r \lesssim 10\%$. Only at the largest lattice constants, where the island-island interactions are weakest, does the manifold break down and the specific energy approach the random value, i.e. that for equal probabilities for each of the 16 vertices.

\begin{figure}[!!t]
\epsfxsize 3.5 in
\epsfbox{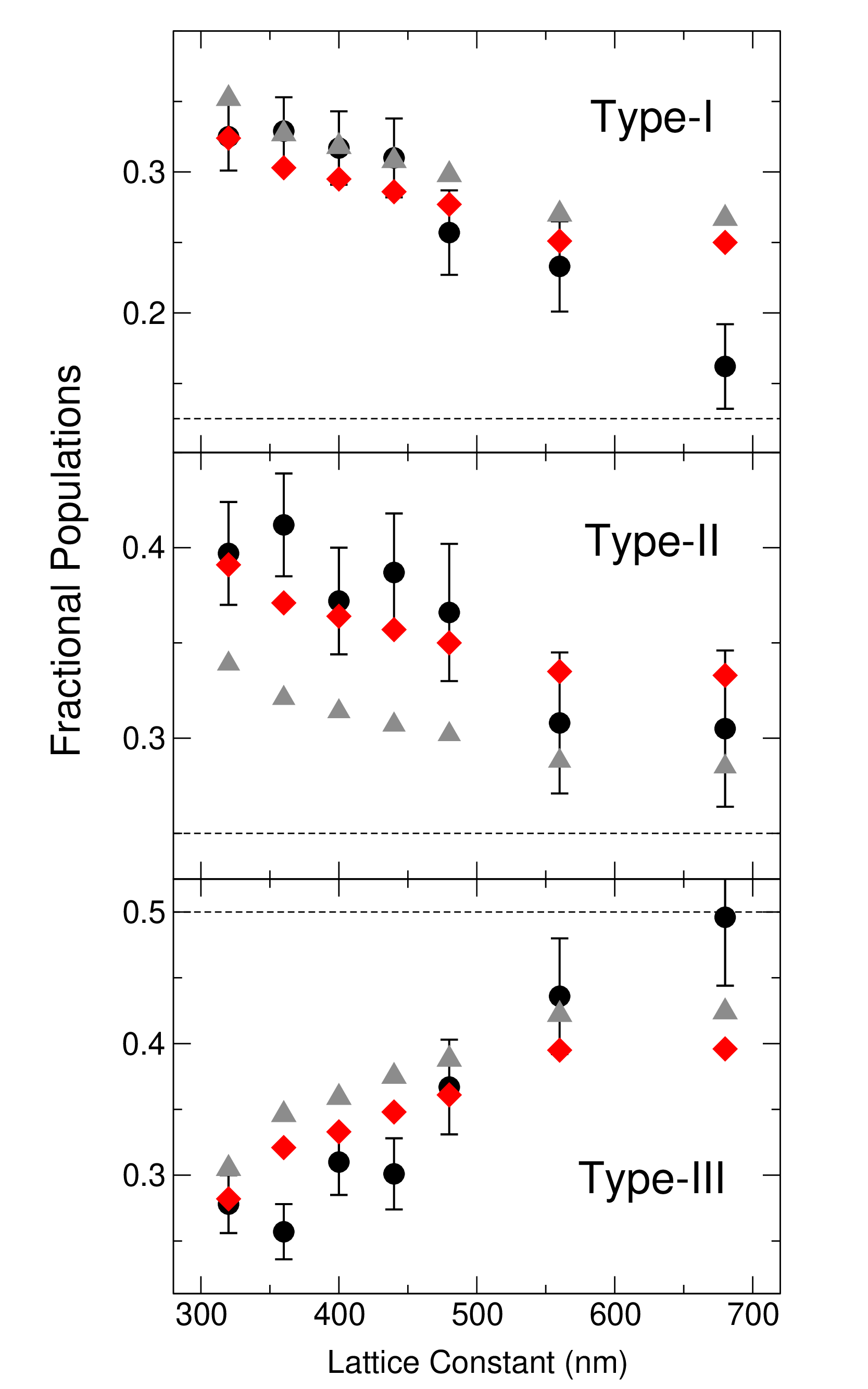}
\caption{Experimental population fractions of vertex types (dots)   compared to the values predicted by maximizing the entropy on the   type-II vertex energy manifold, with (diamonds) and without   (triangles) the background contribution. Dashed lines give the population fraction of a random tiling. Error bars represent statistical counting errors in the populations averaged across sample runs.}
\vspace{-2 mm}
\label{stats}
\end{figure}

To characterize the demagnetized state of this athermal system,  and  to predicts its statistics, we attempt a description as the most likely outcome of demagnetization process.  During demagnetization, the rotating sample is placed in a time varying magnetic field~\cite{protocol}.  At first, the field is large enough to polarize all the islands into a type-II configuration. As the field magnitude drops, it carves defects inside that background. These defects, treated as a non-interacting gas, may appear as types I, II, III or IV, but the background within which they live is purely type-II. We introduce $\rho$, the density of defected vertices; it relates the fractional populations of each vertex type within the defected population, written as
$\nu_{\mathrm{\scriptscriptstyle I}}$,
$\nu_{\mathrm{\scriptscriptstyle II}}$,
$\nu_{\mathrm{\scriptscriptstyle III}}$ and
$\nu_{\mathrm{\scriptscriptstyle IV}}$, 
to the total populations $n_{\mathrm{\scriptscriptstyle I}}$,
$n_{\mathrm{\scriptscriptstyle II}}$, $n_{\mathrm{\scriptscriptstyle
    III}}$, $n_{\mathrm{\scriptscriptstyle IV}}$ via
\begin{equation}
\begin{array}{l} 
 \!\ n_{\mathrm{\scriptscriptstyle I}}=\rho \nu_{\mathrm{\scriptscriptstyle I}} \\
 \!\ n_{\mathrm{\scriptscriptstyle III}}=\rho \nu_{\mathrm{\scriptscriptstyle III}} \\ 
 \!\ n_{\mathrm{\scriptscriptstyle IV}}=\rho \nu_{\mathrm{\scriptscriptstyle IV}}  \\
 \!\ n_{\mathrm{\scriptscriptstyle II}}= (1-\rho) + \rho \nu_{\mathrm{\scriptscriptstyle II}}.
 \end{array} 
\label{n}
\end{equation}
We adopt a vertex-gas approximation~\cite{MC}, wherein each vertex is treated as an independent entity. Thus there are 
\begin{equation}
M=\frac{N!}{ \left(N-D\right)! \prod_{\alpha} N_{\alpha}!}
\label{mult}
\end{equation}
ways to tear  $D= \rho N$ defects in the $N$ vertices of a  polarized tiling, allocated  among the four vertex types with  distribution  $N_{\alpha}=\rho \nu_{\alpha}$, $\alpha=$ I, \dots, IV. We model the demagnetization protocol as a one-step non-equilibrium stochastic process at the vertex level. To obtain the most likely outcome of the vertex populations after demagnetization, we maximize the logarithm of the multiplicity $M$  of Eq.~\ref{mult}, 
\begin{eqnarray}
s = -\left[ \rho \ln \rho +(1-\rho) \ln(1-\rho)\right] + \rho  \sigma,
\label{entropy}
\end{eqnarray}
formally an entropy, under an energy constraint $\overline{E} = E_{\mathrm{\scriptscriptstyle II}}$. The second term of Eq.~\ref{entropy} 
\begin{eqnarray}
  \sigma = & -& \nu_{\mathrm{\scriptscriptstyle I}} \ln 
  \left(\frac{\nu_{\mathrm{\scriptscriptstyle I}}}{2}\right) -\nu_{\mathrm{\scriptscriptstyle II}} \ln \left( \frac{\nu_{\mathrm{\scriptscriptstyle II}}}{4}\right) \nonumber \\
  &-& \nu_{\mathrm{\scriptscriptstyle III}} \ln \left(\frac{\nu_{\mathrm{\scriptscriptstyle III}}}{8}\right)-\nu_{\mathrm{\scriptscriptstyle IV}} \ln \left(\frac{\nu_{\mathrm{\scriptscriptstyle IV}}}{2}\right),
\label{sigma}
\end{eqnarray}
is formally the entropy of an ensemble of vertices $\alpha$, each with probability $\nu_{\alpha}$ and given multiplicity. 
However, if taken as the real entropy  for the actual vertex population $n_{\alpha}$, or $-n_{\mathrm{\scriptscriptstyle I}} \ln \frac{1}{2}
n_{\mathrm{\scriptscriptstyle I}} -n_{\mathrm{\scriptscriptstyle II}}
\ln \frac{1}{4} n_{\mathrm{\scriptscriptstyle II}}
-n_{\mathrm{\scriptscriptstyle III}} \ln \frac{1}{8}
n_{\mathrm{\scriptscriptstyle III}} -n_{\mathrm{\scriptscriptstyle IV}} \ln \frac{1}{2} n_{\mathrm{\scriptscriptstyle IV}}$, and maximized, $\sigma$ would return populations in only modest agreement with the experimental results for type-II, as seen in Fig.~\ref{stats}. As already stressed, demagnetization is not thermal equilibration. Maximization of the ``entropy'' $s$ of Eq.~\ref{entropy} simply returns the most likely outcome of the demagnetization. 

Since background vertices all contribute $E_{\mathrm{\scriptscriptstyle II}}$ to the energy, the energy constraint is relevant only in the maximization of the ``inner entropy''
$\sigma$: $E_{\mathrm{\scriptscriptstyle I}}
\nu_{\mathrm{\scriptscriptstyle I}}+E_{\mathrm{\scriptscriptstyle II}}
\nu_{\mathrm{\scriptscriptstyle II}}+E_{\mathrm{\scriptscriptstyle
    III}} \nu_{\mathrm{\scriptscriptstyle
    III}}+E_{\mathrm{\scriptscriptstyle IV}}
\nu_{\mathrm{\scriptscriptstyle IV}} =E_{\mathrm{\scriptscriptstyle
    II}}$.  The defect population fractions that result are written
$\{\nu^*_{\alpha}\}$.  Next, $s$ is maximized relative to the total defect population $\rho$ with $\sigma$ evaluated at $\{\nu^*_{\alpha}\}$, yielding 
\begin{equation}
\rho^*=\frac{1}{e^{-\sigma^*}+1} 
\label{rho}
\end{equation}
The residual magnetization (as a fraction of the saturation magnetization, which is pure type-II) is bounded above by the background fraction: $M_r \le 1-\rho$~\cite{sqrt2}.

The vertex population fractions obtained by this procedure are compared to the experimental results in Fig.~\ref{stats}. The agreement is good, with no adjustable parameters. This agreement, if taken to verify the model, implies a striking feature in the process of rotational demagnetization: in probabilistic terms, the background state stands on equal footing with each of the defect states. Considering the history of each island's magnet state during demagnetization (in the rotating frame), this suggests that each vertex has a single opportunity to leave the background state and all histories are equally probable.

As the lattice constant increases, the energy difference between types I and II increases faster than that between II and III. Therefore, under the energy constraint, the ratio $n_{\mathrm{\scriptscriptstyle III}}/n_{\mathrm{\scriptscriptstyle I}}$ increases: the defected sample acquires larger multiplicity and $\rho$ must increase; hence type-III increase and type-II decreases.  The fraction of type-IV is too small to make strong statements, although its population increase at larger lattice constants is also predicted by theory. 

The predictions are least accurate at large lattice constant, where island-island interactions are weak and the vertex approximation breaks down, since the system approaches a limit of ideal independent dipoles. At large lattice constants, the system is observed to deviate from the manifold of constant vertex energy (see inset of Fig.~\ref{energy}), approaching the random-tiling values expected for non-interacting islands.  Eqn.~\ref{rho} is therefore expected to not apply in this regime. Nevertheless, it can still predict at least the population of type-II vertices, which are those most affected by $\rho$: assuming that the random tiling fraction holds for the defected vertices (i.e. $\nu_{\mathrm{\scriptscriptstyle II}}=1/4$),  Eqn.~\ref{rho} implies $\rho=16/17$, so we obtain $n_{\mathrm{\scriptscriptstyle II}} = 1-\frac{16}{17} +\frac{16}{17} \frac{1}{4} \simeq 0.294 \dots$, close to the experimental values of $0.305\pm0.013$ for 680 nm and $0.287\pm0.028$ for 880 nm.  These arrays appear to occupy an intermediate regime where the energy constraint has broken down, but not yet the stochastic hypothesis. 

In summary, we have shown that artificial spin ice, despite the absence of thermal fluctuations, can be described within an entropy maximization formalism, resulting in effective thermodynamic behavior for an athermal system. A demagnetized state with an extensive frozen-in residual entropy (similar to that seen in water and spin ice) lives on a sub-manifold of constant vertex energy as the most likely outcome of a minimal stochastic model of the demagnetization process. All nontrivial structure in the microstate can be subsumed into the residual polarized component, with the remainder distributed according a Gibbs-type ensemble. The experimental results are closely matched  with no adjustable parameters. The dynamical origin of this behavior remains to be explained, and may relate to the step size chosen experimentally being the maximal decrement in absolute field that demagnetizes the array. Behavior at smaller step sizes could reveal further details as to the full extent of this regime.

We gratefully acknowledge financial support from the Army Research Office (DAAD19-03-1-0236) and the National Science Foundation MRSEC program (DMR-0213623).


\vspace{-5mm}
\begin{thebibliography}{}
\vspace{-3mm}

\bibitem{Ramirez1} A.~P. Ramirez in {\it Handbook of Magnetic Materials Vol. 13} (ed. K. J. H. Buschow) 423--520 (Elsevier   Science, Amsterdam 2001).

\bibitem{Moessner} R.~Moessner, Can. J. Phys. {\bf 79}, 1283
(2001).

\bibitem{Pauling} L.~C. Pauling, J. Am. Chem. Soc. {\bf 57}, 2680 (1935); {\it The Nature of the Chemical Bond}, Cornell Univ. Press, Ithaca, NY (1945).

\bibitem{Harris} M.~J. Harris, S.~T. Bramwell, D.~F. McMorrow,   T.~Zeiske and K.~W. Godfrey, Phys Rev. Lett. {\bf 79}, 2554
  (1997).

\bibitem{Ramirez2} A.~P. Ramirez, A.~Hayashi, R.~J. Cava,   R.~Siddharthan, and B.~S. Shastry, Nature {\bf 399}, 333
  (1999).

\bibitem{Bramwell} S.~T. Bramwell and M.~J.~P. Gingras, Science {\bf     294}, 1495
(2001).

\bibitem{Wang} R.~F. Wang, C.~Nisoli, R.~S. Freitas, J.~Li, W.~McConville,   B.~J. Cooley, M.~S. Lund, N.~Samarth, C.~Leighton, V.~H. Crespi and   P.~Schiffer, Nature (London) {\bf 439}, 303
(2006).

\bibitem{Moller} G.~M\"oller and R.~Moessner, Phys. Rev. Lett. {\bf 96},   237202 (2006)

\bibitem{Baxter} R.~Baxter, {\it Exactly Solved Models in Statistical Physics} (Academic, New York, 1982).
                                                           
\bibitem{LW} E.~H. Lieb and F.~W. Wu, in {\it Phase Transitions and Critical Phenomena}, vol. 1, C.~Domb and M.~S. Green, Eds. (Academic, London, 1971).

\bibitem{Lieb} E.~H. Lieb, Phys. Rev. Lett. {\bf 18}, 692 (1967).

\bibitem{protocol} 
The field magnitude is reduced from 1280 Oe (well above the coercive field of an island) to 800 Oe in steps of 32 Oe, then to zero primarily in steps of 16 Oe, with a sweep rate of 24000 Oe/second and a hold time of one second at each field value. The sample rotates at 1000 RPM. Increasing the hold time ten-fold does not substantively change the results. The step size of 16 Oe is the maximal such value that successfully demagnetizes the sample. The magnetic moment of a single island, 80 $\times$ 220 $\times$ 25 nm thick, is $\sim 3\times 10^7 \mu_B$. Details are available in R.~F.~Wang, J.~Li, W.~McConville, C.~Nisoli, X.~Ke, J.~W.~Freeland, V.~Rose, M.~Grimsditsch, P.~Lammert, V.~H.~Crespi and P.~Schiffer, J. Appl. Phys. in press.

\bibitem{OtherTypes} 
Calculations using the Object Oriented MicroMagnetic Framework yield 
$E_{\mathrm{\scriptscriptstyle III}} -  E_{\mathrm{\scriptscriptstyle II}}=0.67$
and  
$E_{\mathrm{\scriptscriptstyle IV}} - E_{\mathrm{\scriptscriptstyle II}}=3.4$ 
in units of $E_{\mathrm{\scriptscriptstyle II}} -  E_{\mathrm{\scriptscriptstyle I}}$, 
at a lattice constant of 320 nm (http://math.nist.gov/oommf).

\bibitem{couple}The external field couples most strongly and coherently to the large collective dipole moments associated with   any long-ranged demagnetization fields.

\bibitem{ansatz}
The reason behind the energy ansatz, and whether it arises coincidentally from counteracting kinetic tendencies or coherently from a cooperative process, is still elusive.  Entropy maximization on the vertex gas under the constraint of a fixed specific vertex energy does yield the largest background fraction, and hence the largest possible interaction with external field, when $\overline{E}$ equals the specific vertex energy of the background.
 
\bibitem{MC}Monte Carlo simulations in naive thermal equilibrium provide more general support for a vertex model. For example, vertex energies corresponding to the 320 nm lattice yields populations within a few percent of the vertex gas predictions for temperatures above about $1.3   \left(E_{\mathrm{\scriptscriptstyle II}}-E_{\mathrm{\scriptscriptstyle I}}\right)$, which is   comparable to $T_{\mathrm{eff}}$ for that array.  However, keep in mind that the state described in the main text is not the naive 16-vertex equilibrium state.

\bibitem{sqrt2}$M_r=N^{-1}\sqrt{(M_1^2+M_2^2)/2}$ where $N$ is the number of unit cells and $M_1,~M_2= \pm 1$ are the magnetic moments of the two islands within a unit cell. 

\end{thebibliography}
\end{document}